# The status of the human gene catalogue

Paulo Amaral[1], Silvia Carbonell-Sala[2], Francisco M. De La Vega[3], Tiago Faial[4], Adam Frankish[5], Thomas Gingeras[6], Roderic Guigo[2,7], Jennifer L Harrow[8], Artemis G. Hatzigeorgiou[9], Rory Johnson[10], Terence D. Murphy[11], Mihaela Pertea[12,13], Kim D. Pruitt[11], Shashikant Pujar[11], Hazuki Takahashi[14], Igor Ulitsky[15], Ales Varabyou[11,16], Christine A. Wells[17], Mark Yandell[18], Piero Carninci[13,19,*], and Steven L. Salzberg[11,12,15,20,*]

[1]INSPER Institute of Education and Research, São Paulo, SP, Brasil
[2]Centre for Genomic Regulation (CRG), Dr. Aiguader 88, 08003, Barcelona, Catalonia, Spain
[3]Department of Biomedical Data Science, Stanford University School of Medicine, Stanford, CA; Tempus Labs, Inc., Chicago, IL
[4]Nature Genetics, San Francisco, CA, USA
[5]European Molecular Biology Laboratory, European Bioinformatics Institute, Wellcome Genome Campus, Hinxton, Cambridge CB10 1SD, UK
[6]Department of Functional Genomics, Cold Spring Harbor Laboratory, Cold Spring Harbor, NY
[7]Universitat Pompeu Fabra (UPF), Barcelona, Catalonia, Spain
[8]Centre for Genomics Research, Discovery Sciences, AstraZeneca, Da Vinci Building. Melbourn Science Park, Royston UK SG8 6HB
[9]Universithy of Thessaly, Department of Computer Science and Biomedical Informatics, Lamia, Greece; Hellenic Pasteur Institute, Athens, Greece
[10]School of Biology and Environmental Science, University College Dublin, D04 V1W8 Dublin, Ireland; Conway Institute of Biomedical and Biomolecular Research, University College Dublin, D04 V1W8 Dublin, Ireland; Department of Medical Oncology, Inselspital, Bern University Hospital, University of Bern, 3010 Bern, Switzerland; Department for BioMedical Research, University of Bern, 3008 Bern, Switzerland
[11]National Center for Biotechnology Information, National Library of Medicine, National Institutes of Health, Bethesda, MD 20894, USA
[12]Center for Computational Biology, Johns Hopkins University, Baltimore, MD, USA
[13]Department of Biomedical Engineering, Johns Hopkins University, Baltimore, MD, USA
[14]Laboratory for Transcriptome Technology, RIKEN Center for Integrative Medical Sciences, Yokohama Kanagawa 230-0045 Japan
[15]Department of Immunology and Regenerative Biology; Department of Molecular Neuroscience, Weizmann Institute of Science, Rehovot 76100, Israel
[16]Department of Computer Science, Johns Hopkins University, Baltimore, MD, USA
[17]Stem Cell Systems, Department of Anatomy and Physiology, Faculty of Medicine, Dentistry and Health Sciences, The University of Melbourne, Parkville 3010 Vic Australia
[18]Departent of Human Genetics, Utah Center for Genetic Discovery, University of Utah, Salt Lake City, UT, USA
[19]Human Technopole, via Rita Levi Montalcini 1, Milan 20157 Italy
[20]Department of Biostatistics, Johns Hopkins University, Baltimore, MD, USA
*To whom correspondence should be address: salzberg@jhu.edu, piero.carninci@fht.org

## Abstract

Scientists have been trying to identify all of the genes in the human genome since the initial draft of the genome was published in 2001. Over the intervening years, much progress has been made in identifying protein-coding genes, and the estimated number has shrunk to fewer than 20,000, although the number of distinct protein-coding isoforms has expanded dramatically. The invention of high-throughput RNA sequencing and other technological breakthroughs have led to an explosion in the number of reported non-coding RNA genes, although most of them do not yet have any known function. A combination of recent advances offers a path forward to identifying these functions and towards eventually completing the human gene catalogue. However, much work remains to be done before we have a universal annotation standard that includes all medically significant genes, maintains their relationships with different reference genomes, and describes clinically relevant genetic variants.

**Introduction**

The Human Genome Project (HGP) was launched in 1990 with two central goals: "analyzing the structure of human DNA" and "determining the location of all human genes" [1]. The recent sequencing and assembly of a complete human genome from telomere to telomere[2] accomplished the first of these goals: a complete, gap-free DNA sequence. Achieving the second goal, though, has been far more complicated than originally anticipated, despite a vast increase in our knowledge of the location and function of tens of thousands of human genes. Over time, the task of identifying genes and their functions has been augmented with the goal of identifying their regulatory mechanisms. International efforts have been launched to find all functional elements in the genome[3,4], including genes as well as transcriptional and post-transcriptional regulatory elements.

Early conceptions of the genome treated it as a repository for genes, most of which were thought to encode a single protein-coding transcript[5,6]. Today, though, we know that the picture is different, and that human biology can be influenced by thousands of alternative transcripts and transcribed elements that are not translated into proteins[7,8], and by hundreds of thousands of regulatory elements[3]. Further complicating matters, we now know that many transcribed RNA molecules are further processed into smaller RNA fragments that can have functions different from their parent transcripts.

The purpose of this perspective is to revisit the goals of the HGP in light of our increased understanding of the diversity of functional elements in the human genome. While the genome contains many different features, this perspective will focus on genes. In the sections that follow, we will consider how we can finish specific aspects of human gene annotation in the years to come. These include (1) completing the list of protein-coding genes and all of their isoforms; (2) compiling a complete list of RNA genes of all lengths and varieties; and (3) identifying medically important genes and gene variants, and linking them to specific disorders. For each of these discussions, we will review where we are today, and what remains to be done, and then finally (4) we discuss technology needed to complete the annotation of human genes.[9]

**Protein-coding genes**

Protein-coding genes included in major genome annotation databases–e.g., GENCODE, RefSeq, and CHESS–or captured in reference protein annotation databases such as UniProtKB generally have evidence not just for their translation but also, in many cases, for the function of the protein that they encode[10-14]. Primary evidence can include the direct biochemical or molecular experiments or inference of function recovered from the scientific literature. The direct observation of function of a gene product or that of a close paralog provides confidence in the assignment of function of the gene and its annotation as protein-coding. In addition, the generation of high-quality genome sequences for a large number of vertebrate species, alongside the development of software (such as PhyloCSF++ [15], PhastCons [16], PhyloP [17]) capable of using alignments to identify regions of the genome under purifying selection, as well as direct evidence of translation from mass spectrometry data, increases our confidence in many protein-coding genes.

*Protein-coding gene count*

The annotation of protein-coding genes was the primary focus of the Human Genome Project, after capturing the sequence itself, and while this annotation is still incomplete, the scientific community is approaching a consensus on the identities of these genes. From an initial estimate of 50,000-100,000 genes in the 1980s, the estimated number has dropped steadily, falling to 30,000-40,000 with the initial publication of the human genome[18,19], and then further to ~20-25,000[6,20], 22,000[21], and just under 20,000 today[2], one recent database release suggests as few as ~19,000 (e.g., 19,370 in GENCODE Release 41).

These refinements came about through a variety of advances, including comprehensive manual review[22], improvements in computational annotation methods and analysis, and the generation of ever greater volumes of high-quality experimental transcriptional data. Despite the overall reduction in gene count, novel protein-coding genes continue to be identified, as well as alternative isoforms of known genes.

The MANE collaboration[23] recently published a near-complete dataset containing one isoform for each protein-coding gene for which two of the leading annotation projects, RefSeq and GENCODE, agree completely. MANE 1.0 contains 19,062 gene loci, which covers ~95% of the total protein-coding loci of the major human gene catalogs. This ongoing project offers the promise of providing a definitive answer to the question of how many protein-coding genes we have. An important caveat is that the MANE annotation is provided on the human reference genome known as GRCh38, which still contains gaps, and not on the finished T2T-CHM13 assembly, which was reported to contain 140 additional protein-coding genes[2].

We propose a number of future steps to completing the annotation of protein-coding genes in the human genome:

1. For each protein-coding gene, develop a comprehensive picture of its transcripts and their expression levels in all tissues and cell types available, and determine its conservation in other species.
2. For all proteins that fold into stable structures, determine their 3-dimensional structure and evaluate their stability.
3. Determine all alternative sites of transcription initiation and termination, and record how frequently each site is utilized in normal tissues.
4. Label all reproducible splicing events that lead to non-functional proteins.
5. Catalog and highlight the many exceptional cases where normal rules appear to be violated. These include (a) bicistronic genes, where two distinct protein-coding genes occur on the same transcript; (b) selenoproteins, which use UGA to code for selenocysteine rather than as a stop codon; (c) non-standard splice sites with recognition sites deviating from the most common GT-AG, GC-AG, and AT-AC sites[24]; and (d) extremely short exons, which are often missed or misplaced by current methods.

Although we are nearing consensus on a protein-coding gene set, the precise set of annotated protein isoforms is still in flux [12,25]. Determining this number has been challenging for multiple reasons. First, the determination of isoforms today relies primarily on assembly of RNA-seq data, which in turn relies on having a complete sample of all genes in all cell types, including those prevalent during early development. Efforts such as GTEx[26] have surveyed a large number of tissues, but still only cover a subset of cell types. Projects such as the Human Cell Atlas aim to identify cell-type-specific RNAs for all human cell types, but much work remains. Second, computational methods do not consistently produce the same splice isoforms from large, complex RNA-seq data sets, in part because short-read RNA-seq sequencing is insufficient to unambiguously determine complete splice structures. And third, even for those isoforms that do appear reproducibly in RNA-seq experiments, many may not encode functional proteins.

*Pseudogenes*

Another major challenge, beyond identifying the genes and splice variants themselves, is determining which gene-like elements are pseudogenes. Pseudogenes are sequences that represent defective copies of genes: over 14,000 have been annotated on the human genome. They can be divided into three types: processed (introns removed during retrotransposition), unprocessed (introns retained during duplication), and unitary (pseudogenes without a functioning counterpart in human). Recent evidence using long-read technology suggests that some previously-annotated pseudogenes may in fact be functional [27,28], and other reports indicate that some pseudogenes continue to be translated, although the protein products might not be functional[29].

**Noncoding RNA genes**

Non-coding RNA genes (ncRNAs) include a range of different RNA molecules that are transcribed from DNA, that do not encode proteins, and that provide a function in the cell. A variety of subclasses of ncRNAs have been described, including both long ncRNAs (lncRNAs), defined as RNAs ≥200 nt, and shorter ncRNAs such as microRNAs (miRNAs). We note that although many non-functional RNA sequences might be transcribed in various cells and conditions, our definition will only call them genes if they have a discernable function at the cellular or organismal level. Working out which RNAs are functional, among the many that have been annotated, is one of the major challenges ahead. In the near term, most annotation efforts strive to comprehensively catalogue ncRNA transcripts, regardless of their functional status.

Although annotation strategies that search for conserved protein sequences cannot be used for characterizing ncRNAs, high-throughput RNA-seq experiments have provided an abundant source of evidence for transcription of these genes. Compared to protein-coding RNAs, ncRNAs discovered through RNA-seq appear in relatively low abundance, raising questions about whether they encode functional elements or instead represent transcriptional noise. On the larger question of what ncRNA genes do, many possible functions have been described, including regulating expression of other genes, splicing, chromatin architecture, epigenetic regulation, dysregulation in cancer and other diseases, translation, DNA repair, and more[30-32]. And although tens of thousands of ncRNA transcripts are currently annotated in the human genome, their heterogeneity, poorly understood biology, and other characteristics make the comprehensive discovery of all genes in the ncRNA catalogue an unsolved problem.

A summary of ncRNA gene annotation in current catalogues is shown in Table 1. The two most-widely used are RefSeq and GENCODE, both of which employ human annotators along with large-scale cDNA and RNA sequencing resources[10,33-35] to determine which ncRNA genes to include. In parallel, a variety of consortia and individual research laboratories have provided valuable additional resources, including NONCODE, the FANTOM consortium's CAT resource, LNCipedia, miTranscriptome, CHESS, LncBook, RNAcentral, and others (e.g., see [36]).

The overlap between these annotation databases is relatively low[34], illustrating how far we are from a consensus on the identification of ncRNA genes. This rather fragmented landscape has nonetheless delivered an impressive achievement in charting the enormous variety of noncoding RNA genes.

Table 1: Annotation databases that catalogue long ncRNA genes (figures as of late 2022). Here, "long" refers to loci ≥200nt.

| Resource | LncRNA genes |
| --- | --- |
| RefSeq[10] | 17,948 |
| GENCODE[12] | 19,933 |
| NONCODE[37] | 96,411 |
| FANTOM CAT[38] | 27,919 |
| LNCipedia[39] | 56,946 |
| miTranscriptome[40] | 58,648 |
| CHESS[14] | 17,623 |
| LncBook[41] | 95,243 |

*Other challenges to ncRNA annotation*

A variety of evidence suggests that ncRNA catalogues remain incomplete in a number of ways, and the community is still far from agreement on the true number of ncRNA genes and the true number of transcript isoforms. These issues arise from a variety of sources. First, the transcriptomic datasets from which most ncRNAs are derived originate from a non-exhaustive set of tissues/cell types that are over-represented by adult organs, cell lines and tumors. Rare but important cell types (e.g., tissue stem cells) or difficult-to-access developmental timepoints (e.g., embryonic stages) are poorly represented. This leads to incomplete sampling of existing gene loci and transcript isoforms. Second, the majority of transcriptomic data is produced using oligo-dT reverse transcribed RNA, which largely omits less-studied transcripts such as non-polyA and circular RNAs, although different approaches have been used to circumvent these issues (e.g. [42]). Third, incomplete reverse transcription of cDNA gives rise to transcript models with inaccurate 5' ends, and RNA degradation (which affects major organs at different rates post mortem) can lead to fragmented annotations and incorrect transcription start site (TSS) annotation.

The unique biology of ncRNAs also contributes to the challenges of annotating them. Current evidence indicates that they tend to be expressed at low levels[43], although this might be explained by technical biases in bulk RNA sequencing[44], or in very specific cell types and tissues, leading to relatively infrequent sampling compared to protein-coding RNAs. Their splicing and post-transcriptional processing tends to be as complex as that of protein-coding genes, leading to an ensemble of transcript isoforms that confuses short-read assemblers and human annotators alike[45]. Note that these same features might also be true of non-functional (noisy) transcripts.

Another challenge arises from the dissonance between standard annotation schemas, involving clearly defined, yet arbitrarily defined genes and transcripts, with the messy biological reality of ncRNA transcriptional units. Conventionally, genes are defined as the union of all overlapping transcripts at a locus, and neighboring genes are separated by a clear gap. These definitions worked well in the past. However, with the advent of deep and comprehensive long-read RNA sequencing, annotations are approaching a point at which read-through transcription events will begin to unite nearly all pairs of neighboring genes. Following classical gene definitions, the result could be a single "super gene" on each chromosome[46,47], which is clearly not a useful abstraction.

*Functional annotation*

One of the biggest challenges in ncRNA annotation relates to adding functional labels. For protein-coding genes, we have a rich amount of prior functional evidence, in addition to powerful computational methods for predicting gene function based on primary sequence. For example, DNA-binding transcription factors or membrane-bound receptors can often be predicted from translated amino acid sequences. In contrast, we know little about the vast majority of ncRNAs, and have no validated means of predicting function from sequence. Thus, one near-term goal for annotation of ncRNA genes will be describing the different types of evidence supporting them (e.g., tissue-specific expression levels), even though their function might remain unknown.

To date, many ncRNAs have been assigned names or biotypes that imply some function[48]; in particular, ncRNAs are often named after a nearby or overlapping protein-coding gene. For example, FAS-AS1 is an anti-sense (AS) transcript whose name reflects its overlap with the protein-coding gene FAS. This may lead to confusion amongst users, because the lncRNAs in question may not have a function related to that of the neighboring protein-coding gene.

**Health and medical annotation**

A key application of human gene annotation is its use in diagnosing and treating genetic disease. Over five thousand genes and many thousands of variants of those genes have been associated with single gene disorders and disease risk, as catalogued in OMIM[49]. For example, the BRCA Exchange database (https://brcaexchange.org/) currently lists over 34,000 variants in the *BRCA1* gene alone, of which 2,228 are labeled as pathogenic[50].

When assessing variant pathogenicity in a clinical setting, the completeness and accuracy of gene and transcript models is essential. The impacts of variants as determined by programs such as PolyPhen[51], Revel[51], and Variant Effect Predictor (VEP[52]) depend on the predicted open reading frames of transcripts. Further, designs of oligonucleotide baits and PCR primers used in targeted capture sequencing for clinical diagnostic assays depend on the correct annotations of exon boundaries. Even when whole-genome sequencing (WGS) is used for diagnosis, clinicians do not consider unannotated exons as candidates for interpretation.

Flaws in annotation can lead to serious errors in the clinic. Among many examples that might be cited, one case of a false negative diagnosis was caused by missing exons in a transcript of *CDKL5*, in a proband with seizures who was ultimately diagnosed by WGS after reannotation detected the missing exons[53,54]. Another striking case led to a new diagnosis of Dravet syndrome after reannotation of an isoform of *SCN1A* revealed that the original annotation was missing a "poison" exon. In that case, the patient had splicing variants leading to expression of the nonfunctional isoform[54].

*The need for a clinical standard*

Currently, clinical laboratories often operate on the GRCh37 (hg19) human assembly and use RefSeq transcripts as a reference for well-known disease-linked genes, typically relying on reports from the literature. When the literature is unclear, laboratories tend to choose a transcript using simple criteria such as length or first appearance in annotation databases. This practice has two major problems: first, not all RefSeq transcripts map perfectly onto the GRCh37 human reference; and second, the chosen transcript might not reflect the properties needed for clinical diagnosis or the most representative transcript.

A universal annotation standard to describe clinical variants should satisfy several goals:

1. Every human gene of medical significance must be included, with a single canonical isoform in most cases. If possible, this isoform should be selected to represent all annotated exons, to ensure that the resulting ORF yields a stable protein product, and to ensure that documented medically relevant protein positions are preserved; e.g. *BRAF* V600E.
2. In cases where two or more isoforms have well-documented clinical significance, or when a single isoform cannot represent all annotated exons for a gene, those isoforms should be included as well.
3. All isoforms of each gene should map at 100% identity to GRCh38, and their relationships with RefSeq and GENCODE transcript IDs should be maintained.
4. Mappings of these transcripts to other assemblies including GRCh37 and T2T-CHM13 should be created and maintained, allowing applications developers to include the same genes and transcripts regardless of which human reference genome they use.

Within MANE, transcript isoforms that satisfy criteria (1) above are referred to as "MANE Select," and for cases identified per criteria (2) above, additional isoforms labeled "MANE Plus Clinical" are included. MANE Plus Clinical transcripts are assigned in consultation with clinical experts, and this set is expected to grow as new variants are discovered and documented. We believe that MANE provides a logical starting point for a new, related effort to create clinically important reference annotations of noncoding RNAs and regulatory elements, at least for those that have been associated with genomic variants linked to disease risk [55-57]. Clinical interpretation and reporting will also benefit from more databases mapping their contents to the MANE standard. Databases also need to use standardized descriptions of genetic variants [58] to ensure unambiguous mapping to reference genomes.

*Consistent annotation across multiple reference genomes*

The hg19 (GRCh37) genome was replaced in 2014 by GRCh38. Although both reference genomes are still in use, they differ in many ways: their coordinates are different, some genes are missing from the older version, gene symbols have changed, and many genes have different exon-intron structures. Even for genes that are unchanged across the two releases, there is no standard way to translate coordinates between genomes without creating artifacts.

The advent of a truly complete human genome sequence, T2T-CHM13, promises to provide much more stability in gene coordinates.[2] Looking ahead, we are likely to have many reference genomes for different human sub-populations. We already have annotated reference genomes for Ashkenazi[59], Puerto Rican[60], and Han Chinese individuals[61], and many more are likely to be produced, as well as a "pan-genome" representing all populations. Ultimately, we need a gene-centric way of referring to the same gene, and the same mutations, on any human reference genome.

**Technology to finish the human gene catalogue**

Finishing the human gene catalogue will require innovative new technologies to address the challenges ahead, such as resolving the functional relationships between gene products in a diversity of tissues, cells, and developmental stages. Here we touch on a few technologies that are available now or that may be available soon to solve these problems.

*Matched long-read sequencing and proteomic analysis of gene products.* Genome-wide measurements of when and where specific isoforms are expressed are currently needed. Measuring gene expression within tissues and at single-cell resolution has already revealed many coordinated patterns of gene expression in cells and tissues[62]. However, cell-specific splicing estimates from these studies remain problematic[63], and the number of splicing events is likely underestimated[64].

RNA-seq analysis at the isoform level currently relies on differential expression of exons within a gene[63], which is highly dependent on the method of library construction and on sequencing depth[65]. Even when expression levels are measured accurately, the relative abundance of a transcript does not correlate perfectly with translation[66]. Ribosomal profiling is a powerful method for measuring the translation of protein-coding isoforms, and it can validate engagement with the translational machinery for many predicted alternate isoforms. Interestingly, transcript analysis from ribosomal scanning and translation complexes at polysome fractions predicts large numbers of unannotated small ORFs[67,68], and these need further exploration to determine if they represent valid functional genes. These predictions might be resolved through full-length sequencing, preferably directly from RNA molecules, coupled with further ribosomal profiling or other methods for detecting translation.

While sequencing with single molecule technologies (e.g., Oxford Nanopore Technologies (ONT) and Pacific Biosciences (PacBio)) is capable of providing full-length direct RNA and cDNA sequences, relatively few experiments to date have used these technologies to survey the RNA landscape from each human cell type. Other confounding issues concern sequencing the poly-A terminating RNAs, using ONT oligo-dT reverse transcription primers and oligo-dT linker ligation[69]. Strategies are being developed to specifically capture total RNA that will use RNA ligases to add a primeable sequence at the true 3' ends of all RNA transcripts. Another approach uses artificial poly(U) tailing to add a primeable sequence to both capped RNAs and non-capped RNAs[70,71]. Information on RNA modification, which can be measured from ONT direct RNA sequencing data, will likely provide a powerful new type of functional annotation.

Validation of protein-coding isoforms will ultimately require protein detection. Meta-analyses of proteomic data rely heavily on the quality of the transcriptome reference to identify peptides mapping to putative isoforms. However, when coupled with new long-read technologies, dual proteome-transcriptome assemblies are finding evidence of higher isoform diversity than predicted from a representative transcript approach, by resolving peptide fragments that would otherwise fail to map unambiguously to a gene or single isoform. In one recent study, 30% of the gene products identified using a dual PacBio-mass spectrometry approach were distinct isoforms from the same locus, which included thousands of examples where the alternate isoform was not measurable using mass spectrometry alone[72]. We anticipate that soon, progress in long-read technologies will produce more reliable maps of full-length transcript isoforms, quantifiable isoform switching, and isoform dosage at the resolution of individual cells.

*Methods to capture low-expressed transcripts.* Capture sequencing has recently been adapted to target specific RNAs, in order to provide higher sequencing coverage for selected regions of the genome using short- and long-read RNA sequencing in a high-throughput manner[33,73]. This is particularly useful to enrich for RNA from ultra-low input samples[74] and from genes expressed at very low levels. The use of capture technologies, together with recent increases in the throughput of long-read sequencing platforms, could enormously benefit the study of low-expressed transcripts, particularly lncRNAs, which in turn may be vital for the study of gene regulation in both normal and diseased cells.

**Conclusions**

Over 20 years after the original publication of the human genome, the number of protein-coding genes is stabilizing around 19,500, although the number of isoforms of these genes is still a subject of intensive study and discussion. The completion of a human genome sequence itself offers the opportunity to map these genes onto a stable, finished sequence and converge to a final number in the next few years. Greater standardization of gene and isoform annotation will improve our ability to apply this knowledge in a clinical setting.

In contrast, noncoding RNA genes, particularly lncRNAs, are at an earlier stage of understanding, and are still increasing in number, with current catalogs containing 17,000-20,000 lncRNAs or more. New technologies offer promising avenues to refine this catalog, although a complete functional characterization of lncRNAs is likely many years away. The steady decline in the number of protein-coding genes over the last 20 years makes it only natural to ask if lncRNA numbers may follow a similar trend, as our knowledge of RNA biology and technologies improve.

Finally, we note that even with a complete gene annotation of a finished genome, we will have only one example of the human gene catalogue, one that will not apply to all humans. It is likely that many healthy individuals have more or fewer copies of some genes, and future efforts to survey the diversity of the human population will be an important step towards achieving a more complete view of the gene content of our genome.

**Acknowledgements**
Thanks to the Banbury Center at Cold Spring Harbor Laboratory and the Cold Spring Harbor Laboratory Corporate Sponsor Program for supporting a workshop that all authors of this work attended. This work was supported in part by: U.S. National Institutes of Health (NIH) under grants R01-HG006677 (MP, SLS, AV), R01-MH123567 (MP, SLS), R35-GM130151 (SLS), U41-HG007234 (AF), and U24-HG007234 (RG, SC); the Wellcome Trust under grant WT222155/Z/20/Z (AF); the European Molecular Biology Laboratory (AF); the U.S. National Science Foundation under grant DBI-1759518 (MP); Science Foundation Ireland through Future Research Leaders award 18/FRL/6194 and the Irish Research Council through Consolidator Laureate award (IRCLA/2022/2500) (RJ); the National Center for Biotechnology Information of the National Library of Medicine, NIH (TDM, KDP, SP); National Health and Medical Research Council (NHMRC) APP1186371 (CAW); Center for Genomic Medicine at the University of Utah Health, and the H.A. & Edna Benning Foundation (MY); Spanish Ministry of Science and Innovation to the EMBL partnership, Centro de Excelencia Severo Ochoa and CERCA Programme / Generalitat de Catalunya (RG, SC); RIKEN Center for Integrative Medical Sciences (PC, HT); and Human Technopole (PC).